\documentclass[12pt]{article}
\usepackage{epsf}
\usepackage{amsmath}

\usepackage{graphics}
\usepackage{cite}

\begin{titlepage}

\begin{document}

\hfill July 2019

\begin{center}

{\bf \Large Holographic Principle, Cosmological Constant and Cyclic Cosmology \\
}
\vspace{2.5cm}
{\bf Claudio Corian\`{o} and Paul H. Frampton\\  }
\vspace{0.5cm}
{\it Dipartimento di Matematica e Fisica "Ennio De Giorgi",\\ 
Universit\`{a} del Salento and INFN-Lecce,\\ Via Arnesano, 73100 Lecce, Italy\footnote{claudio.coriano@le.infn.it, paul.h.frampton@gmail.com}
}

\vspace{1.0in}

\begin{abstract}
\noindent
The holographic principle provides a deep insight into quantum gravity and
resolves the fine-tuning crisis concerning the cosmological constant. 
Holographic dark energy introduces new
ultra-violet (UV) and infra-red (IR) cutoffs into quantum gravity which are 
necessarily strongly related. The equation of state for dark energy
$\omega = p/\rho$ is discussed from the holographic point of view.
The phantom option of $\omega < -1$
is resurrected, as in an earlier cyclic cosmology. Such a cyclic model
can, however, equally use the cosmological constant with $\omega = -1$.

\end{abstract}

\end{center}

\bigskip
\bigskip
\bigskip
\bigskip

\noindent
Published in Mod. Phys. Lett. A34, 1950355. (2019)

\end{titlepage}

\section{Introduction} 

\noindent
The 1998 discovery of accelerating cosmic expansion gave rise
to a theoretical explanation by a small non-zero cosmological
constant(CC). At first the magnitude of the CC seemed 
exceedingly small when compared with the Planck scale.

\bigskip

\noindent
Five years earlier, 't Hooft \cite{Hooft} had made the stunning suggestion
that the number of degrees of freedom for gravity in (3+1)
spacetime is the same as the number of degrees of freedom
for quantum field theory in (2+1) spacetime. It did not take long
\cite{CKN}  to try to connect
these two observations by pointing out that the usual calculation of the
cosmological constant in quantum field theory had not
taken account of 't Hooft's drastic reduction in the number
of gravitational degrees of freedom.
The initial work did not lead to the correct equation of state $\omega=p/\rho$
for the dark energy\cite{Hsu}. This led to an interesting modification \cite{Li,NO2005,NOS2019}
which provides the jumping off point here.

\bigskip

\noindent
There are some puzzles remaining in this approach \cite{KimLeeKim} but since it makes
dramatic progress towards the magnitude of the CC it is worth asking whether it can gain
some traction in handling the equally vexing questions surrounding
cyclic cosmology which confronts the Tolman Entropy
Conundrum(TEC). This will be discussed in the present paper.

\bigskip

\noindent
As we shall discuss in the next section, the holographic principle 
dictates that we take care to choose cutoffs such that they do
not allow states which lie inside their own Schwarzschild radius.
This imposes strong constraints which dramatically modify
how we approach calculations in quantum gravity.

\section{IR and UV Cutoffs}

\noindent
The na\"{i}ve estimate of the cosmological constant
from quantum field theory (QFT) uses the vacuum energy
from the 0-point function and results in

\begin{equation}
\Lambda_{QFT} \sim \int^{M_{Planck}} d^3k \sqrt{k^2+m^2} \sim M_{Planck}^4
\label{CC}
\end{equation}
so that, using the reduced Planck mass $M_{Planck} \sim 10^{18}$GeV,
one estimates
\begin{equation}
\Lambda_{QFT} \sim 10^{72} (GeV)^4 \equiv 10^{108} (eV)^4
\label{CC2}
\end{equation}
to be compared with the observational value
\begin{equation}
\Lambda_{Obs} \sim 10^{-12} (eV)^4 
\label{CC3}
\end{equation}
displaying the 120 orders of magnitude discrepancy between
theory, $\Lambda_{QFT}$, and experiment, $\Lambda_{Obs}$.
It is fair to say that before the advent \cite{Hooft} of the holographic principle,
this discrepancy was simply described as the largest error in theoretical physics
and defied any explanation. 
Employing the holographic principle, however, the expression in
Eq.(\ref{CC}) for $\Lambda_{QFT}$ includes a significant overestimate
of the number of degrees of freedom. The point is that the UV cutoff is really
much less than $M_{Planck}$. Let us denote this
ultraviolet cutoff for gravity by $M_{UV}$ and the infra-red cutoff 
by $M_{IR} = L_{IR}^{-1}$ where $L_{IR}$ is the size of the system.

\bigskip

\noindent
The UV cutoff in the gravitational sector in Eq.(\ref{CC}) must
be reduced to $10^{-30} M_{Planck}$ if the calculation is to
be consistent with observation. That such a dramatic reduction
is feasible is testament to the power of the holographic principle.

\bigskip

\noindent
According to the holographic principle, the volume $L_{IR}^3$ 
occupied by the effective field theory describing gravity must
satisfy that its entropy is less than that of a black hole of
radius $L_{IR}$. This requires the inequality
\begin{equation}
L_{IR}^3 M_{UV}^3 < L_{IR}^2 M_{Planck}^2
\label{holo1}
\end{equation}
 which implies a scaling law
 \begin{equation}
 L_{IR} \propto \left( \frac{1}{M_{UV}^3} \right)
 \label{scaling}
 \end{equation}
 
 \bigskip
 
 \noindent
 Even Eqs. (\ref{scaling}) is insufficiently strong to avoid
 disallowed states whose Schwarzschild radius exceeds $L_{IR}$. To see
 this, consider the effective field theory at a temperature satisfying
 \begin{equation}
 M_{IR} \ll T < M_{UV}
 \label{T}
 \end{equation}
 In the volume $L_{IR}^3$ the thermal energy $E$ and entropy $S$ are given
 by $E=L_{IR}^3 T^4$ and $S=L_{IR}^3 T^3$ respectively. If we saturate the inequality
 of Eq.(\ref{holo1}) we find a system with Schwarzshild radius $R_S$ given
 by
 \begin{equation}
 R_S = \frac{E}{M_{Planck}^2}= L_{IR} (L_{IR} M_{Planck})^{\frac{2}{3}} \gg L_{IR}
 \label{Schwarzschild}
 \end{equation}
 which confirms that Eq.(\ref{scaling}) is insufficiently strong to exclude states
 whose Schwarzschild radius exceeds the size of the box.
 
 \bigskip
 
 \noindent
 To exclude all states which lie within their Scwarzschild radius requires that one
 impose the stronger inequality
 \begin{equation}
 L_{IR}^3 M_{UV}^4 \leq L_{IR} M_{Planck}^2
 \label{holo2}
 \end{equation}
 which implies that
  \begin{equation}
 L_{IR} \propto \left( \frac{1}{M_{UV}^2} \right)
 \label{scaling2}
 \end{equation}
 which is much stronger than Eq.(\ref{scaling}). When Eq.(\ref{holo2})
 is saturated the maximum entropy $S_{max}$ falls short of the 
 black hole entropy $S_{BH}$ by
 \begin{equation}
 S_{max} = S_{BH}^{\frac{3}{4}}
\label{Smax}
\end{equation}
The significant difference between $S_{max}$ and $S_{BH}$
resides in states which are not describable by conventional  
quantum field theory. Therefore, we do not insist on Eq.(\ref{Smax})
but allow $S_{max}=S_{BH}$.

\bigskip

\noindent
However, there remains a fatal flaw in the discussion so far as the alert
reader may have noticed. The point it that by using the constraint in
Eq.(\ref{scaling2}) the cosmological constant $\Lambda$ develops
a dependence on the FLRW scale factor $a(t)$ of the form
\begin{equation}
\Lambda \propto \left( \frac{1}{a(t)^3} \right)
\label{badL}
\end{equation}
which uses the fact that $L_{IR} \propto a(t)^{\frac{3}{2}}$.

\bigskip

\noindent
But the scaling of Eq.(\ref{badL}) means that the dark energy and matter
terms on the right-hand-side of the Friedmann expansion equation
behave similarly and that therefore the dark energy has equation
of state $\omega_{DE}=p/\rho = 0$ corresponding to pressureless dust
rather than $\omega_{DE} < -\frac{1}{3}$ as necessary for accelerated
expansion.

\bigskip

\noindent
The holographic approach to dark energy thus appeared doomed
until the appearance of paper\cite{Li} which made an interesting
proposal of how to proceed more successfully. The idea is to replace
the radius of the  visible universe by the future event horizon 
$L_{IR}=R_h$ as the infrared cutoff
given by
\begin{equation}
R_h = a \int_t^{\infty} \frac{dt}{a} = a \int _a^{\infty} \frac{da}{H a^2}
\label{Rh}
\end{equation}
This future event horizon is the boundary of the volume a fixed observer may
eventually observe.

\bigskip

\noindent
Writing
\begin{equation}
\rho_{DE} = 3 c^2 M_{Planck}^2 L_{IR}^{-2}
\label{rhoDE}
\end{equation}
and assuming dominance by dark energy in the Friedmann expansion equation
\begin{equation}
H^2 = \frac{1}{3 M_{Planck}^2} \rho_{DE}
\label{Friedmann}
\end{equation}
we find that
\begin{equation}
R_h H = c
\label{RhH}
\end{equation}
and consistency requires with a new normalisation that
\begin{equation}
\frac{1}{H} = \left( \frac{\alpha}{ca} \right) a^{\frac{1}{c}}
\label{1overH}
\end{equation}
which means that the equation of state $\omega=p/\rho$ 
satisfies
\begin{equation}
- 3 (1 + \omega) = -2 (1 - \frac{2}{c})
\end{equation}
which means that
\begin{equation}
\omega = - \frac{1}{3} - \frac{2}{3c}
\label{omegaDE}
\end{equation}
in which $c > 0$.

\bigskip

\noindent
From Eq.(\ref{omegaDE}) we see that $\omega < - \frac{1}{3}$
as required for accelerated expansion and that $\omega = -1$
when $c=1$, corresponding to a cosmological constant.

\bigskip

\noindent
Fits to the observational data tend to favour $c \leq 1$ 
corresponding to $\omega \leq -1$, although $c > 1, \omega > -1$
cannot yet be excluded.

\section{Cosmological Constant}

\noindent
Observationally the magnitude of the cosmological constant $\Lambda$
is approximately $\Lambda \sim +10^{-12} eV^4$ and its equation of state is
$\omega = p/\rho \simeq -1$, quite closely.

\bigskip

\noindent
From the previous section, setting $c = 1$, we have
\begin{equation}
\Lambda = 3 M_{Planck}^2 R_h^{-2}
\label{Lambda2}
\end{equation}
and using $M_{Planck} = 10^{27}$ eV and $R_h^{-1} = 10^{-33}$ eV
this gives immediately a result $\Lambda=+10^{-12} eV^4$, consistent with observation.
Compared to Eq.(\ref{CC}) in the Introduction we notice that
the holographic principle has decreased the UV cutoff by 30
orders of magnitude and hence the CC, which goes like
the UV cutoff to the fourth power, by 120 orders of
magnitude. This provides vindication of the radical proposal in \cite{Hooft}
about quantum gravity.

\bigskip

\noindent
Let us discuss the equation of state $\omega$ given by Eq.(\ref{omegaDE})
in the precious section. With $c=+1$ we find $\omega=-1$.
Consistent with observational data, the parameter may instead be, 
for example, $c=0.986 < 1$ which corresponds to
$\omega=-1.01$ and is characteristic not of a cosmological constant
but of phantom dark energy. This small-seeming change
drastically changes the fate of the universe.
Both $\omega=-1$ exactly and $\omega=-1.01$ will be interesting
cases for our ensuing discussion about cyclic cosmology.

\section{Cyclic Cosmology}

\noindent
There is an undeniable attraction to the idea of a cyclic universe
which goes an infinite number of times through an 
\begin{equation}
{\rm expansion} \longrightarrow {\rm turnaround}
\longrightarrow {\rm contraction} \longrightarrow {\rm bounce} \longrightarrow {\rm expansion} 
\end{equation}
 process. 
In the earliest days of theoretical
cosmology most of the leading theorists (De Sitter, Einstein, Friedmann,
Lema\^{i}tre, Tolman) at some point favoured such a theory, primarily
to avoid the initial singularity present in the Friedmann expansion
equation.

\bigskip

\noindent
However, considerations of entropy and the second law of thermodynamics
led Tolman\cite{Tolman1,Tolman2} in 1931 to a no-go theorem about
cyclic cosmology, often called the TEC(=Tolman Entropy Conundrum).
Simply put, if the entropy continuously increases as required by the
second law, each cycle becomes larger and longer. Correspondingly, 
in the past the cycles were smaller and
shorter and therefore must have at some finite past time
originated from an initial singularity.

\bigskip

\noindent
Entropy of the universe enters our considerations not only because
off the TEC but also because of the necessity of exceptionally low
entropy at the beginning of the present expansion era. Why should
the universe be in such a homogenous uniform state at the start?
Cyclic cosmology should address also this second entropy issue
which is not explained in, for example, inflationary theory.

\bigskip

\noindent
What we have in mind is an infinitely cyclic theory with an infinite
past. The infinite past raises interesting mathematical issues which were
addressed in the 2009 preprint\cite{PHFpreprint}. The cyclic model
we shall discuss has, at present, an infinite number of universes
forming an infiniverse. This will remain the case for the infinite
future. What is more subtle is the infinite past where according to
\cite{PHFpreprint} there are two possibilities: (A) there was always
be an infinite number of universes; (B) by using the set theory
idea of {\it absence of precedent} it can begin, an infinite time
in the past, with a finite number of universes, possibly only one.

\bigskip

\noindent
Of course, what was unknown to Tolman and to all other theoretical
cosmologists until the end of the twentieth century is the
dark energy which drives the observed accelerated cosmological
expansion. This provides alternatives to prior thinking, providing novel ways
to get rid of the entropy of our universe, for example at the
turnaround from expansion to contraction. One important issue
is to provide observational tests for a given model.
In \cite{DEexpt} it was
shown, based on conservative and plausible assumptions,
that to be sensitive to any effects of dark energy an experiment
must be at least the size of a galaxy. It is therefore discomfiting
to read a recent paper \cite{ATLASexpt} looking for dark 
energy at the LHC.
Although the LHC is the largest scientific apparatus ever
constructed, nevertheless it falls short of the size of the Milky Way by 
many orders of magnitude. Although \cite{DEexpt}
emphasises $\omega < -1$ the argument therein applies equally
to $\omega = -1$.

\bigskip

\noindent
The equation of state $\omega = p/\rho$, where $p$ is pressure
and $\rho$ is density, plays an important r\"{o}le although not as
important as first thought when emphasis was (mis)placed on the
phantom possibility $\omega < -1$ which can lead to a big 
rip\cite{Caldwell}, a little rip \cite{FLS} or one of its variants
\cite{FLNOS,FLS2,FL}. As we shall discuss, the proposals
for a cyclic cosmology survive in the case of $\omega=-1$
which is the equation of state for the cosmological constant.

\bigskip

\noindent
{\bf Underlying the CBE hypothesis is the idea that as the
universe undergoes accelerated expansion, especially faster
then exponential as in the Big Rip, it will be torn apart into
causally disconnected regions which continue to expand
until the turnaround. Almost all of these regions will be
empty of matter and contract to a bounce after
turnaround. The vanishingly small number of regions
containing matter will bounce prematurely and are
failed universes. The successful empty universes, of which
ours is one, contract, or come back, adiabatically with
vanishing entropy until the bounce. The CBE hypothesis
is a speculation and has not been justified mathematically.}

\bigskip

\noindent
In the model of \cite{BF} the method of evading the TEC 
was based on the Come Back Empty (CBE) idea.
The CBE hypothesis is that our contracting universe contains
no matter, including no black holes, only entropy-free dark matter
and hence contract adiabatically with zero entropy. The almost
vanishing number of other universes which do contain matter
and / or black holes will be failed universes because they will
prematurely bounce after the turnaround from expansion to
contraction.

\bigskip

\noindent
To discuss cyclicity, the holographic model with $c = 0.986, \omega = -1.01$
from the previous section is closest to the original discussion of \cite{BF}.
We first note that the time from the present time $t_0$ to the
big rip at $t=t_{rip}$ is\cite{FT}
\begin{equation}
t_{rip} - t_0 = \left( \frac{11Gy}{-\omega_{DE} - 1} \right) \simeq 1.1 Ty
\label{trip}
\end{equation}
so that to one-digit accuracy we can say $t_{rip} = t_T = 1.0Ty$ where
$t_T$ is the time of turnaround from expansion to contraction which
is only a fraction of a second before $t_{rip}$.

\bigskip

\noindent
At $t_T$ the universe divides into a very large number $N$ of causally
disconnected patches, almost all of which are empty of matter
and of black holes. The vanishingly small number of causal
patches which do contain quarks and leptons and black holes
will necessarily fail to contract all the way to a normal bounce
because the matter will proliferate and cause a premature
bounce. The successful universes, of which ours is one,
can contract to a bounce a fraction of a second before the
would-be big bang.

\bigskip

\noindent
The {\it total} entropy of the infiniverse always increases consistent
with the second law of thermodynamics, but at turnaround
the entropy of {\it our} universe drops very close to zero and remains
nearly vanishing until the bounce, thereby explaining why the
entropy at the beginning of the next expansion era is
very low.

\bigskip

\noindent
With the bounce at time $t=0$, during the expansion phase
 $0 < t < t_T$ the
scale factor $a(t)$ satisfies the Friedmann
expansion equation
\begin{equation}
\left( \frac{\dot{a}(t)}{a(t)} \right)^2 =
\frac{8\pi G}{3} \left[ \frac{ (\rho_{\Lambda})_0}{a(t)^{3(\omega_{\Lambda}+1)}}
+ \frac{(\rho_m)_0}{a(t)^3} + \frac{ (\rho_r)_0}{a(t)^4} - \frac{\rho_{TOT}(t)^2}{\rho_c} \right]
\label{unhatFriedmann}
\end{equation}

\bigskip

\noindent
After turnaround at $t = t_T$, the scale factor deflates to
$\hat{a}(t_T) = f a(t_T)$ where $f < 10^{-28} \propto N^{-\frac{1}{3}}$
and a fraction $(1-f)^3$ of the entropy is jettisoned at turnaround.
During the contraction from $t=t_T$ to the bounce at $t=t_B$
the reduced scale factor satisfies a Friedmann contraction
formula
\begin{equation}
\left( \frac{\dot{\hat{a}}(t)}{\hat{a}(t)} \right)^2 =
\frac{8\pi G}{3} \left[ \frac{ (\hat{\rho}_{\Lambda})_0}{\hat{a}(t)^{3(\omega_{\Lambda}+1)}}
+ \frac{ (\hat{\rho}_r)_0}{\hat{a}(t)^4} - \frac{\hat{\rho}_{TOT}(t)^2}{\hat{\rho}_c} \right]
\label{hatFriedmann}
\end{equation}
with
\begin{equation}
\hat{\rho}_i(t) = \frac{(\rho_i)_0 f^{3(\omega_i+1)}}{\hat{a}(t)^{3(\omega_i+1)}}
= \frac{(\hat{\rho}_i)_0}{\hat{a}(t)^{3(\omega_i+1)}}
\label{hatrho}
\end{equation}
and $\hat{\rho}_m = 0$ because of the CBE hypothesis.
The CBE assumption was critically examined
in \cite{BF2} where it was confirmed that after
turnaround the universe contains at most one photon.
As the contraction progresses, spatial flatness
is rapidly approached as an attractor point of Eq.(\ref{hatFriedmann}).

\bigskip

\noindent
We must comment on the different case $c=+1, \omega=-1$
of the previous section which is the cosmological constant.  
Note that the discussion of the holographic principle
no longer requires a Big Rip.
At first sight this is very different because
there is no would-be Big Rip. However, it is straightforward
to show that a turnaround and
bounce can occur in a similar way by employing
a right-hand-side to the Friedmann expansion equation
containing $\left( \rho_{DE} - \frac{\rho_{DE}^2}{\rho_c} \right)$
as can be justified by higher
dimensional brane models {\it e.g.} \cite{Lisa1,Lisa2}.

\bigskip

\noindent
{\bf In the paper \cite{BF} it was assumed that the cosmic
expansion was faster than exponential and heading for a would-be
Big Rip in order that the required break-up into causally-disconnected
regions be sufficient. The new result of the present paper
which uses the holographic principle
is that the CBE hypothesis in \cite{BF} can be assumed also
for an expansion which is exactly exponential, as is the case
for a cosmological constant. In one important sense, this is better because
it avoids the issues of negative energy densities which can occur
in the Big Rip}.

\bigskip

\noindent
An analysis in \cite{BFM} showed that CBE is feasible for
any $\omega_{\Lambda} < -2$, which includes all the
values of interest, since it allows the number $N$ of
causal patches at turnaround be sufficient 
to satisfy the CBE constraint. Another study \cite{Frampton}
showed that scale invariant density perturbations in
the radiation field are provided during contraction
which re-enter the horizon after the bounce. Finally,
the typical time elapse before the turnaround was
estimated\cite{Frampton2} by a new matching condition
method which reassuringly resulted in a value of $t_T$
consistent with that estimated in Eq.(\ref{trip}) above.

\section{Discussion}

\noindent
As we have seen, the holographic principle strongly
modifies previous calculations in quantum gravity
made without the benefit of its knowledge. Most
striking is the diminution of the magnitude of
the cosmological constant by 120 orders of
magnitude, thereby ending that mystery.

\bigskip

\noindent
This requires a decrease of 30 orders of magnitude
in the UV cutoff $M_{UV}$ for the gravity sector. The IR cutoff
scales like $L_{IR} \propto M_{UV}^{-2}$ and the
correct proportionality constant involves the future
event horizon in order to obtain the correct equation
of state $\omega_{DE} \simeq -1$. This choice for the
IR cutoff is somewhat counterintuitive because it
uses information from the future, or at least assumes
that the accelerated expansion will continue.

\bigskip

\noindent
With respect to cyclic cosmology, the holographic dark
energy permits an equation of state $\omega = -1$
or slightly more negative {\it e.g.} $\omega = -1.01$.
The latter case was emphasised in \cite{BF}.
However, the former case can equally underly
an infinitely cyclic model by appealing to a brane
term in the Friedmann expansion equation.

\end{document}